\documentclass{article}
\usepackage{epsf}

\begin{document}


\pagestyle{empty}

\renewcommand{\thefootnote}{\fnsymbol{footnote}}
                                                  
\begin{flushright}
{\small
SLAC--PUB--8213\\
July 1999\\}
\end{flushright}
                
\begin{center}
{\bf\Large
An Improved Direct Measurement of Leptonic Coupling Asymmetries
with Polarized $Z^0$'s \footnote{Work supported in part by the
Department of Energy contract  DE--AC03--76SF00515.}}

\bigskip
The SLD Collaboration$^{**}$
%
\vskip 1truecm
\end{center}

\vspace{2.5cm}

\begin{abstract}

We report new direct measurements of the $Z^0$-lepton coupling 
asymmetry parameters $A_e$, $A_{\mu}$ and $A_{\tau}$, 
with polarized $Z^0$'s collected by the SLD detector 
at the SLAC Linear Collider.
The parameters are extracted from the measurement of 
the left-right-forward-backward asymmetries for each lepton species.
The 1996, 1997 and 1998 SLD runs are included in this analysis
and combined with published data from the 1993-95 runs.
Preliminary results are
$A_e = 0.1558 \pm 0.0064$, $A_{\mu}=0.137 \pm 0.016 $ and
$A_{\tau}=0.142 \pm 0.016$.
If lepton universality is assumed, 
a combined asymmetry parameter $A_{l} = 0.1523  \pm 0.0057$ results.
This translates into an effective weak mixing angle
$\sin^2\theta^{eff}_W = 0.23085 \pm 0.00073$ at the $Z^0$ resonance.

\end{abstract}

\vspace{2cm}

\begin{center}

{\sl Paper Contributed to 
the XIXth International Symposium 
     on Lepton Photon Interactions, August 9-14 1999, Stanford,  USA.}

\end{center}
\vfill

\normalsize

\pagebreak
\pagestyle{plain}

\pagebreak

%
\newpage

\section*{INTRODUCTION}

The structure of the parity violation in the electroweak interaction can
be probed directly in the production and decay of polarized $Z^0$ bosons.
The parity violations of all three leptonic states are 
characterized by the $Z^0$-lepton coupling asymmetry parameters;
$A_e$, $A_{\mu}$ and $A_{\tau}$. 
The standard model assumes lepton universality, 
so that all three species of leptonic asymmetry parameters
are expected to be identical and directly related to 
the effective weak mixing angle, $\sin^2\theta^{eff}_W$.
Measurements of leptonic asymmetry parameters 
at the $Z^0$ resonance provide an important test of lepton universality
and the weak mixing angle~\cite{Erler:1998ig}.

We report new results on direct measurements of the asymmetry parameters
$A_e$, $A_{\mu}$ and $A_{\tau}$ using leptonic $Z^0$ decays.
The measurements are based on the
data collected by the SLD experiment at the SLAC Linear Collider (SLC).
The SLC produces polarized $Z^0$ bosons
in $e^+e^-$ collisions using a polarized electron beam.
The polarization allows us to form the left-right cross section asymmetry
to extract the initial-state asymmetry parameter $A_e$. 
It also enables us to extract the final-state asymmetry parameter
for lepton $l$, $A_l$, 
directly using the polarized forward-backward asymmetry. 
Experiments at the $Z^0$ resonance
without beam polarization~\cite{Abbaneo:1999ub} 
have measured the product of initial- and
final-state asymmetry parameter, $A_e\cdot A_l$.
Those same experiments have also measured the tau 
polarization~\cite{Abbaneo:1999ub} 
which yields $A_e$ and $A_{\tau}$ separately.
The SLC beam polarization enables us to
present the only existing direct measurement of $A_\mu$.
The polarized asymmetries yield the statistical enhancement 
on the final-state asymmetry parameter by a factor of about 25 
compared to the unpolarized forward-backward asymmetry. 
In this report, we use
the data recorded in 1996-98 at the SLD with the upgraded vertex detector.
The obtained results are combined with 
earlier published results~\cite{Abe:1997xm}. 

There are two principle goals of this study. 
One is to test lepton universality by comparing 
the three asymmetry parameters. 
The other purpose is to complement the weak-mixing-angle result
from the left-right cross section asymmetry 
using the hadronic event sample~\cite{Abe:1997nj}
and to add additional precision to the determination of 
the weak mixing angle.

 
\section*{THE SLC AND THE SLD}

This analysis relies on the Compton polarimeter, 
tracking by the vertex detector and 
the central drift chamber (CDC), and the liquid argon calorimeter (LAC). 
Details about the SLC, the polarized electron source and the measurements of
the electron-beam polarization with the Compton polarimeter, 
can be found in Refs.~\cite{Woods:1996ph} and \cite{Woods:1996nz}.
A full description of the SLD and its performance have also been described 
in detail elsewhere~\cite{unknown:1984rp}.
Only the details most relevant to this analysis are mentioned here.

In the previous measurements~\cite{Abe:1997xm}, 
the analysis was restricted in the polar-angle range of $|\cos\theta|<0.7$ 
because the trigger efficiency
for muon-pair events and tracking efficiency fall off beyond 
$|\cos\theta|=0.7$.
The upgraded vertex detector and new additional trigger system improved 
the acceptance.
The upgraded vertex detector (VXD3)~\cite{Abe:1997bu},
a pixel-based CCD vertex detector, was installed in 1996.
The VXD3 consists of 3 layers which enable a self-tracking capability 
independent of the CDC and provides 3-layer and 2-layer
coverage out to $|\cos\theta|=0.85$ and 0.90, respectively.
The self-tracking capability and wide acceptance of VXD3 give significant
improvement in
solid-angle coverage because high precision VXD3-hit vectors in 3-D are
powerful additions to the global track finding capability.
The detailed implementation of this new strategy to recover deficiencies
in track finding with the CDC alone is already developed, and working
well on recent SLD reconstruction data~\cite{VERTEX99byT.Abe}.
The new additional trigger for lepton-pair events is the WIC Muon Trigger
(WMT).
The purpose of the WMT is to trigger muon-pair events passing through 
the endcaps.
The WMT uses data from the endcap Warm Iron Calorimeter (WIC) 
which consists of inner and outer sections.
The WMT requires straight back-to-back tracks in the endcap WIC
passing through the interaction point.
In order to increase the efficiency of the WMT, 
only one of back-to-back inner or back-to-back outer tracks are required.
Angular coverage of the WMT is $0.68<|\cos\theta|<0.95$ with reasonable 
trigger efficiency covering the lack of leptonic trigger region 
in the previous analysis.

\section*{THEORY}

\subsection*{$A_{LR}$ and $\widetilde{A}_{FB}^l$  }

Polarization-dependent asymmetries are easily computed from the tree-level
differential cross section for the dominant process 
$e^-_{L,R} + e^+ \rightarrow Z^0 \rightarrow l^- + l^+$ at $Z^0$ resonance,
where $l$ represents either a $\mu$- or a $\tau$-lepton. 
The differential cross section is expressed as follows:
\begin{equation}
{{d\sigma}\over{d\cos\theta}}\propto
\left(1-PA_e\right)\left(1+\cos^2\theta\right)+
2\left(A_e-P\right)A_l\cos\theta , \label{Eq:dsigma}
\end{equation}
where $\cos\theta$ is the angle of the outgoing lepton ($l^-$) 
with respect to the electron-beam direction.
Photon exchange terms and, if final-state leptons are electrons,
$t$-channel contributions~\cite{cite:gterm} have to be taken in to account.
The leptonic asymmetry parameters which refer to the initial- and final-state
lepton appear in this expression as $A_e$ and $A_l$, respectively. 
Note that the first term, symmetric
in $\cos\theta$, exhibits initial-state coupling to the electron by its
dependence on $A_e$. 
The second term, asymmetric in $\cos\theta$, is mostly influenced by $A_l$. 
$P$ is the signed longitudinal polarization of the electron beam 
in the convention that left-handed bunches have negative 
sign~\cite{cite:polsign}.

The relationships between the asymmetry parameters and between vector and
axial-vector, or left-right couplings, are given as follows:
\begin{equation}
A_l = {{2{g^l_V}{g^l_A}}\over{{g^l_V}^2 + {g^l_A}^2}}
= {{{g^l_L}^2 - {g^l_R}^2}\over{{g^l_L}^2 + {g^l_R}^2}}. \label{Eq:aldef}
\end{equation}
where $g^l_L = g^l_V + g^l_A$ and $g^l_R = g^l_V - g^l_A$.
The Standard Model relates the weak mixing angle to the couplings by the
expressions 
$g^l_V = -{1 \over 2} + 2\sin^2\theta^{eff}_W$
and 
$g^l_A = -{1 \over 2}$.

Simple asymmetries can be used to extract $A_l$ from data;
the left-right asymmetry and the left-right-forward-backward asymmetry.
The left- and right-handed cross sections are obtained by
integrating Eq.~(\ref{Eq:dsigma}) over all $\cos\theta$ giving $\sigma^l_L$ or
$\sigma^l_R$ for left- and right-handed electron beams, respectively. 
(For convenience, we drop the superscript in the following discussions 
since the meaning of the expressions will be clear enough in context.) 
Parity violation causes $\sigma_L$ and $\sigma_R$ to be different. 
Hence, we define the left-right cross section asymmetry, $A_{LR}$,
\begin{equation}
A_{LR} = {1\over |P|}\ {{\sigma_L - \sigma_R}\over{\sigma_L + \sigma_R}}. 
	\label{Eq:alrdef}
\end{equation}
Four cross sections are obtained by integrating forward (F) and backward (B)
hemispheres separately, along with left- and right-handed polarization. 
Here forward (backward) means $\cos\theta > 0$ ($\cos\theta < 0$).
Based on these four possibilities, we define the polarized forward-backward
asymmetry, $\widetilde{A}_{FB}^l$, as follows:
\begin{equation}
\widetilde{A}_{FB}^l
={{(\sigma_{LF}-\sigma_{LB})-(\sigma_{RF}-\sigma_{RB})}\over
{(\sigma_{LF}+\sigma_{LB})+(\sigma_{RF}+\sigma_{RB})}} .\label{Eq:apfb}
\end{equation}

\subsection*{Leptonic Asymmetry Parameters: $A_e$, $A_{\mu}$ and $A_{\tau}$}

With equal luminosities for left- and right-handed electron beams, the cross
sections in Eq.~(\ref{Eq:alrdef}) may be replaced with the numbers of
events: $N_L$ and $N_R$. After integrating Eq.~(\ref{Eq:dsigma})
over all angles to get expressions for $N_L$ and $N_R$ in terms of $P$, 
$A_e$ and $A_l$, and after substituting  in Eq.~(\ref{Eq:alrdef})
for both signs of polarization, what remains is given by
\begin{equation}
A_e = A_{LR}.
\end{equation}
In a similar fashion, integrating over forward or backward hemispheres, and
substituting both signs of polarization in Eq.~(\ref{Eq:apfb}), 
gives the expression
\begin{equation}
A_l = ({\widetilde{A}_{FB}^l}/|P|)(1+{x^2_{max}\over3})/x_{max}, 
\end{equation}
where $x_{max} = \cos\theta_{max}$ is the maximum
polar angle accepted by the lepton-event trigger and reconstruction
efficiencies.

The leptonic asymmetry parameters are particularly potent ways 
to measure the weak mixing angle precisely 
because $A_l$ is expressed as follows:
\begin{equation}
A_l = \frac{2\left( 1-4\sin^2 \theta_W^{eff} \right)}
	{1+\left( 1-4\sin^2 \theta_W^{eff} \right)^2}
\end{equation}
making $A_l$ very sensitive to the weak mixing angle:
\begin{equation}
\frac{dA_l}{d\sin^2 \theta_W^{eff}} \simeq -7.9 .
\end{equation}
%

\subsection*{The Maximum Likelihood Method }

The essence of the measurement is expressed in
Eqs. (\ref{Eq:alrdef}) and (\ref{Eq:apfb}), but instead of simply
counting events we perform a maximum likelihood fit, event by event, 
to incorporate the contributions of all the terms in the cross section and to
include the effect of initial state radiation.
The likelihood function for muon- and tau-pair events is defined as follows:
\begin{equation}
{\cal L}(A_e,A_l,x)=
\int ds^\prime 
H(s,s^\prime)
\left(
Z(s^\prime,A_e,A_l,x)+Z\gamma(s^\prime,A_e,A_l,x)+\gamma(s^\prime,x)
\right) ,\label{Eq:likelihood function(MIZA1)}
\end{equation}
where $A_e$ and $A_{\mu}$ are free parameters.
$A_e$ and $A_\mu$ ($A_\tau$) are determined simultaneously with 
the muon-pair (tau-pair) events.
The integration over $s^\prime$ is done with the program
MIZA~\cite{Martinez:1991ta} to take into account the initial-state
radiation from two times the beam energy $\sqrt{s}$ to the
invariant mass of the propagator $\sqrt{s^\prime}$
described by the radiator function $H(s,s^\prime)$. 
The spread in the beam energy has a negligible effect.
The maximum likelihood fit is less sensitive to detector
acceptance as a function of polar angle than the counting method,
and has more statistical power.
$Z(...), \gamma(...)$ and Z$\gamma(...)$ are the tree-level differential
cross sections for Z exchange, photon exchange and their interference.
The integration is performed before the fit
to obtain the coefficients $f_Z, f_{Z\gamma}$ and $f_\gamma$,
and the likelihood function becomes
\begin{equation}
{\cal L}(A_e,A_l,x)=
f_Z\cdot Z(A_e,A_l,x)+f_{Z\gamma}\cdot Z\gamma(A_e,A_l,x)+
f_\gamma\cdot\gamma(x). \label{Eq:likelihood function(MIZA2)}
\end{equation}
These coefficients give the relative sizes of the three terms at
the SLC center-of-mass energy. 

As for the electron final state,
it includes both $s$-channel and $t$-channel $Z^0$ and photon exchanges
which gives four amplitudes and ten cross section terms.
All ten terms are energy-dependent.
We define a maximum likelihood function for electron-pair events
by modifying Eqs.~(\ref{Eq:likelihood function(MIZA1)}) and 
(\ref{Eq:likelihood function(MIZA2)})
including all ten terms.
The integration over $s^{\prime}$ is performed with 
DMIBA~\cite{Martinez:1992wy} to obtain the coefficients
to give the relative size of the ten terms. 

\section*{ANALYSIS}

\subsection*{Data Sample}

This study includes the data obtained during the 1996 and 1997-98 SLD runs. 
Results are combined with published analyses from data taken during the 1993
SLD run and the 1994-95 run. 
The 1996 data set consisted of about 50,000 $Z^0$'s with
about $77\%$ polarization. 
The 1997-98 data sample contains about 340,000 $Z^0$'s.
The beam polarization 
for the 1997-98 runs averaged about 73\%. 
The data were recorded at a mean center-of-mass energy of 
91.26 GeV and 91.23 GeV during 96-97 and 98 runs, 
respectively~\cite{ZSCAN}.
The branching ratio $Z^0 \rightarrow l^+l^- = 3.4\%$ so
that the total branching ratio into all three lepton species 
combined is about 10\%.

\subsection*{Event Selection} 

Leptonic $Z^0$ decays are characterized by their low multiplicity and high
momentum charged tracks. 
Muons and electrons are particularly distinctive as
they emerge back-to-back with little curvature from the primary interaction
vertex, and tau pairs form two tightly collimated cones directed in
well-defined opposite hemispheres. 
Lepton-pair candidates are chosen on the
basis of the momentum of the charged tracks as well as from energy deposited in
the calorimeter. 
The criteria used for the event selection give a high
efficiency for finding the signal events while the backgrounds remain
sufficiently low as to be almost entirely negligible. 

\vskip 1truecm

\leftline{\bf The pre-selection:}
\begin{itemize} 
\item Lepton-pair candidates are initially selected by requiring the charged
multiplicity between two and eight charged tracks to reduce background
from hadronic $Z^0$ decays. 
\item The product of the sums of the charges of the
tracks in each hemisphere must be -1. 
This insures a correct determination of
the sign of the scattering angle. 
\item Requiring that at least one track have momentum greater than 1 GeV
reduces two-photon background while leaving candidate
events with a high efficiency.
\end{itemize}
After the pre-selection, additional selection criteria are applied. 

\vskip 1truecm

\leftline{\bf The Bhabha selection:}
\begin{itemize}
\item A single additional cut effectively selects $e^+e^-$ final states.
We require the sum of the energies associated with the two highest 
momentum tracks in the event must be greater than 45 GeV as measured 
in the calorimeter.
\end{itemize}

\vskip 1truecm

\leftline{\bf The muon-pair selection:}
\begin{itemize}
\item Muon final state selection starts by demanding that the invariant 
mass of the event, based on charged tracks, be greater than 70 GeV.  
Tau final states usually fail this selection. 
\item Since muons deposit little energy as they traverse
the calorimeters, we require also that the largest energy recorded in the
calorimeter by a charged track in each hemisphere be greater than
zero and less than 10 GeV. 
Electron pairs are removed by this requirement.
\end{itemize}

\vskip 1truecm

\leftline{\bf The tau-pair selection:}
\begin{itemize}
\item Tau selection requires that the largest calorimeter energy associated 
with a charged track in each hemisphere is less than 27.5 GeV and 20.0 GeV for
the magnitude of $\cos\theta$ less than or greater than 0.7, respectively, to
distinguish them from $e^+e^-$ pairs.  
\item We take the complement of the muon event
mass cut and require the event mass to be less than 70 GeV. 
\item At least one track
must have momentum above 3 GeV to reduce backgrounds from two-photon events. 
\item We define the event acollinearity from the vector sums of the momenta 
of the tracks in the separate hemispheres, 
and the angle between the resultant 
momentum vectors must be greater than 160 degrees.
This also removes two-photon events.
\item Finally, the invariant mass of charged tracks in each hemisphere is 
required to be less than 1.8 GeV to further suppress hadronic backgrounds.
\end{itemize}

The results from the event selections are summarized in 
Table~\ref{Table:background-and-efficiency}. 
Each event is assigned a polar production angle based on the thrust axis
defined by the charged tracks.
Our published results based on the 1993-95 data were restricted to 
the polar-angle range $\left|\cos\theta\right|<0.7$ because of 
the lepton trigger and the tracking acceptance~\cite{Abe:1997xm}.
In the 1996-98 data sets, we can use a wider polar-angle range
$\left|\cos\theta\right|<0.8$ for the analysis.

\begin{table}
\begin{center}
\caption{\label{Table:background-and-efficiency} 
Background and efficiencies for 1996-98 data.}
\begin{tabular}{lccc} \hline
Sample & Background & Efficiency for  & \# of Events    \\
       &               & $\left|\cos\theta\right|<0.8$ &  \\
\hline
$e^+e^-$ 	& 1.2\% $\tau^+\tau^-$ & 87.3\% &   14803  \\
\hline
$\mu^+\mu^-$	& 0.2\% $\tau^+\tau^-$ & 85.5\% &   11867  \\ \hline
$\tau^+\tau^-$	& 0.7\% $e^+e^-$       &        &         \\
                & 2\% $\mu^+\mu^-$     & 78.1\% &   11266  \\
                & 1.7\% 2$\gamma$      &        &         \\
                & 0.8\% hadrons        &        &    \\ \hline
\end{tabular}
\end{center}
\end{table}

\begin{figure}[h]       
\centerline{\epsfxsize 3.5 truein \epsfbox{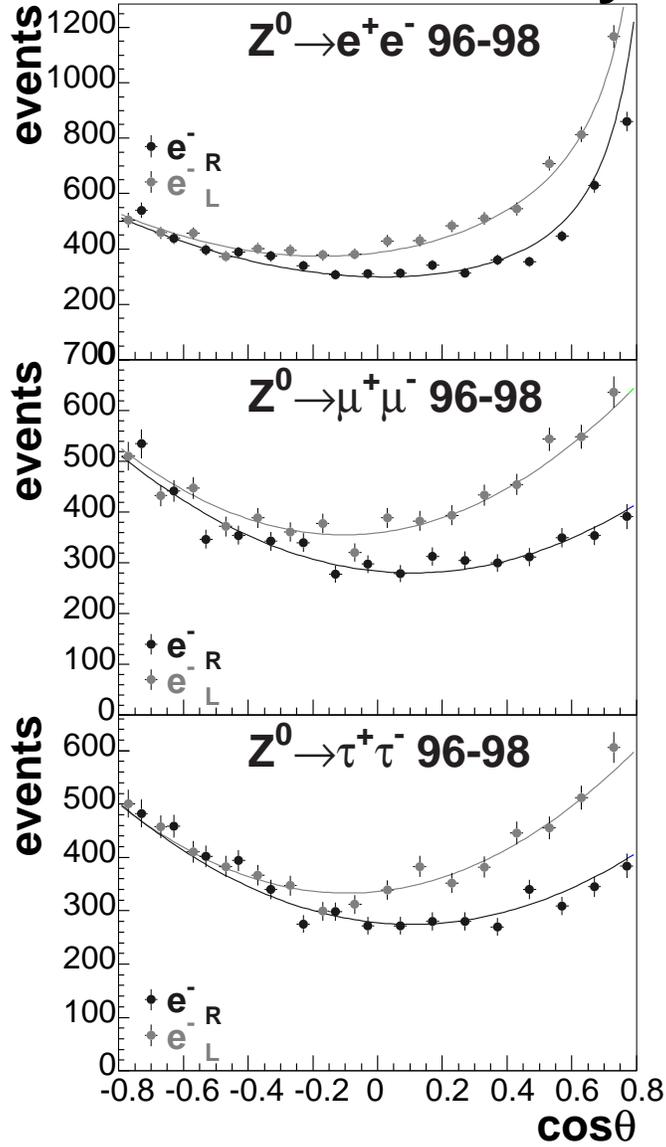}}
\caption{Polar-angle distributions for $Z^0$ decays to $e$, $\mu$ and $\tau$ 
pairs for 1996-98 SLD runs.
The open(filled) circles are for left(right)-handed electron polarization.
The data are corrected for $\left|\cos\theta\right|>0.7$ where the detection
efficiency drops with increasing $\left|\cos\theta\right|$. 
For the Bhabha events, the forward-backward asymmetry has the same sign 
for both polarizations, while for the muon- and tau-pair events,
the forward-backward asymmetry changes sign with polarization.
}
\label{Fig:cost}
\end{figure}

Polar-angle distributions for electron-, muon- and tau-pair final states
from the 1996 through 1998 data sets are shown in Fig.~\ref{Fig:cost}. 
The left-right cross section asymmetries and forward-backward angular 
distribution
asymmetries are clearly seen. The acceptance in $\cos\theta$ out to $\pm 0.7$
is uniform, but falls off at larger $\cos\theta$. Since the data are plotted
out to $\cos\theta$ $\pm 0.8$, it was necessary to correct the data
for the acceptance efficiency in order that the fitted curve could be compared
with the data.  
Similar fits were done for the 1993 and 1994-95 data sets. 
In all cases the curves fit the data well.

\subsection*{Systematic Effects and Corrections to Asymmetry Parameters}

The maximum likelihood procedure gives an excellent first estimate of the
asymmetry parameters and the statistical error on each parameter. 
However there are several systematic effects which can bias the result:
\begin{itemize}
\item	Uncertainty in beam polarization;
\item	Background;
\item	Uncertainty in beam energy; and
\item	V-A structure in $\tau$ decay.
\end{itemize}
We must estimate the systematic errors on these effects. 
We discuss these effects in this section and summarize in
Table~\ref{Table:correction} and Table~\ref{Table:Systematic-Error}. 

\vskip 1truecm

\leftline{\bf Effect of polarization asymmetries:}

Asymmetry measurements at SLD rely critically on the time-dependent 
polarization.
SLD has three detectors to measure the polarization, 
Cherenkov detector (CKV)~\cite{Woods:1996nz},
Polarized Gamma Counter (PGC)~\cite{Field:1998sh}
and Quartz Fiber Calorimeter (QFC)~\cite{Berridge:1997pw}.
Due to beamstrahlung backgrounds produced during luminosity running,
only the CKV detector can make polarization measurements during beam 
collisions.
Hence it is the primary detector and the most carefully analyzed.
Dedicated electron-only runs are used to compare electron polarization
measurements between the CKV, PGC and QFC detectors.
The PGC and QFC results are consistent with the CKV result at the level of
0.5\%.
Details on the polarization measurements are discussed in 
Ref.~\cite{cite:polmeasurement}.
The preliminary estimates of the error on the polarization are given by
$\delta P/P = 0.67\%$ and 1.08\% for 1996 and 1997-98, 
respectively~\cite{cite:finalpol}.

\vskip 1truecm

\leftline{\bf Effect of Backgrounds:}

Muon-pair samples are almost background free but tau-pair candidates are
contaminated by electron-pairs, two-photon and hadronic events. 
A small percentage of tau-pairs are identified as electron-pairs. 
Beam-gas and cosmic ray backgrounds have been estimated and found negligible. 
Estimates of backgrounds are given in 
Table~\ref{Table:background-and-efficiency}. 
These estimates have been derived from
detailed Monte Carlo simulations as well as from studying the effect of cuts in
background-rich samples of real data. 
Tau pairs are the only non-negligible backgrounds 
in the electron- and muon-pair samples. 
The tau-pairs background in the muon-pair sample is negligible 
since the world-averaged measurements say $A_{\mu}$ and $A_{\tau}$
are consistent within $13 \cdot 10^{-3}$ and the effect would be smaller
than $5\times10^{-5}$.
For the same reason, the muon-pair background in the tau-pair sample can be 
neglected.
The $t$-channel electron-pair background, the two-photon background and the
hadronic background cause small corrections to $A_{\tau}$. 

We estimate how the backgrounds discussed above affect each asymmetry 
parameter by creating an ensemble of fast simulations. 
The simulation-data sets are generated from the
same formula for the cross-sections used to fit the real data. 
Trial backgrounds are then superimposed on each data sets,
where the shape of the background has been obtained from the shape of
the data that form the particular background.
Each background is normalized relative to the signal according to
detailed Monte Carlo estimates. 
The effect of each background on each asymmetry
parameter is determined from the differences between the fitted
parameter values before and after inclusion of the backgrounds.
The net corrections due to backgrounds and their uncertainties are given in 
Table~\ref{Table:correction}.

\begin{table}
\begin{center}
\caption{\label{Table:correction} 
Systematic corrections to $A_e$ and $A_{\tau}$ from tau and electron
final states for 1996-98 data.
Corrections to $A_{\mu}$ can be neglected.}
\begin{tabular}{lccc} \hline
Final State & Background & $\delta A_e$ ($\times10^{-4}$) & 
$\delta A_{\tau}$ ($\times10^{-4}$)     \\    \hline
$e^+e^-$        & $\tau^+\tau^-$       & $4\pm4$    &  --    \\
\hline
$\tau^+\tau^-$  & $e^+e^-$             &  --        & $-2\pm2$  \\
                & two photon           & $25\pm25$  & $18\pm18$ \\
                & hadron               &  --        & $13\pm13$ \\
                & V-A                  &  --        &$-130\pm29$ \\
{$\tau^+\tau^-$ Totals }   &           & $25\pm25$  &$-101\pm37$  \\
\hline
\end{tabular}
\end{center}
\end{table}

\vskip 1truecm

\leftline{\bf Effect of uncertainty of center of mass energy:}

The calculation of the maximum likelihood function depends on the average
beam energy $\sqrt{s}$ since the coefficients in the likelihood functions
(see Eq.~(\ref{Eq:likelihood function(MIZA2)})) will depend on the 
center-of-mass energy.
During the 1998 run, a $Z^0$-peak scan~\cite{ZSCAN} was performed to provide 
a calibration of the SLD energy spectrometer.
It shows that the spectrometer measurements had a small bias and that SLD has
been running slightly below the $Z^0$ resonance.
Hence we redetermine the coefficients in 
Eq.~(\ref{Eq:likelihood function(MIZA2)}) for 1998 data to correct the effect.
The uncertainty due to a 
$\pm 1\sigma(\sim 50\mathrm{MeV})$ ~\cite{cite:UncertaintyofECM}
variation of the center-of-mass 
energy is estimated by
computing them for the peak energy as well as for the $1\sigma$ variation.

\vskip 1truecm

\leftline{\bf V-A structure in tau decays:}

The largest systematic effect for the tau analysis, indicated in 
Table~\ref{Table:correction}, comes
about because we measure not the taus themselves, but their decay products. 
The longitudinal spin projections of the two taus from $Z^0$ decay are 100\%
anti-correlated: one will be left-handed and the other right-handed. 
So, given the V-A structure of tau decay~\cite{Tsai:1971vv}, 
the decay products from  the $\tau^+$ and the $\tau^-$ 
from a particular $Z^0$ decay will 
take their energies from the same set of spectra.  
For example, if both taus decay to $\pi\nu$, then both pions
will generally be low in energy (in the case of a left handed $\tau^-$ and
right handed $\tau^+$) or both will be generally higher in energy. 
The effect is strong at SLD because the high beam polarization induces 
very high tau polarization as a function of polar-production angle. 
And, most importantly, the sign of the polarization is basically 
opposite for left- and right-handed polarized beam events. 
So a cut on event mass causes polar-angle dependence in
selection efficiency for taus which has the opposite effect for taus from
events produced with the left- and right-handed polarized electron beam. 
Taking all tau decay modes into account, using Monte Carlo simulation, 
we find an overall shift of $+0.0130 \pm0.0029$ on $A_{\tau}$
(where the uncertainty is mostly from Monte Carlo statistics and 
the value extracted from the fit must be reduced by this amount). 
$A_e$ is not affected since the overall relative efficiencies for 
left-beam and right-beam events are not changed much
(only the polar angle dependence of the efficiencies are changed). 

\vskip 1truecm


The above-mentioned systematic effects are non-negligible, 
although small compared with current statistical errors. 
Other potential corrections are discussed below. 
Their effect on the asymmetry parameters is deemed negligible for the
current measurements. 

\vskip 1truecm

\leftline{\bf Effect of detector asymmetries:}

Since there will generally be no bias in the fit as long the efficiency 
is symmetric in $\cos\theta$, there will be a problem only 
if the efficiency for detecting positive tracks is different
from that of negative tracks. 
We estimate this effect by examining the relative numbers of opposite 
sign back-to-back tracks with positive-positive and negative-negative pairs. 
The latter will occur whenever one of the two back-to-back tracks in 
a two-pronged event has a wrong sign of measured charge. 
Double charge mismeasurement is less likely.
The correction for biases due to charge mismeasurement is found to be 
negligible.

\vskip 1truecm

\leftline{\bf Final state thrust angle resolution:}

We have also studied the effect of uncertainty in the thrust axis by
smearing the directions of outgoing tracks. 
Final state QED radiation can affect the determination of the track angle 
particularly for electrons, although we find the angle to be well-determined 
in that case as well. 
The result depends somewhat on how final pairs are selected but this
source of correction is also deemed negligible from our studies. 

\vskip 1truecm

\leftline{\bf Summary of systematic errors:}

Table~\ref{Table:Systematic-Error} summarizes
the systematic errors on the asymmetry parameters due to the
contributing factors discussed above. The superscript on each parameter
indicates the lepton species from which that particular parameter was
determined. For example, $A_e^{\mu}$ refers to the estimate of $A_e$ obtained
through the dependence expressed in Eq.~(\ref{Eq:dsigma}) by analyzing the muon
pairs.
\begin{table}[h]
\begin{center}
\caption{\label{Table:Systematic-Error} 
Summary of systematic uncertainties in units of $10^{-4}$.}
\begin{tabular}{lccccc} \hline
Source & $A_e^e$ & $A_e^{\mu}$ & $A_e^{\tau}$ & $A_{\mu}^{\mu}$ &
$A_{\tau}^{\tau}$   \\ \hline
Polarization    	& 16 & 16 & 16 & 16 & 16 \\
Backgrounds     	&  4 & -- & 25 & -- & 22 \\
Radiative Correction	& 30 &  3 &  3 &  4 &  3 \\
V-A             	& -- & -- & -- & -- & 29 \\ \hline
{\bf Totals }   	& 34 & 16 & 30 & 16 & 40 \\ \hline
\end{tabular}
\end{center}
\end{table}

\section*{RESULTS}

Preliminary results from fits to the 1996-98 data are summarized
below:
\begin{eqnarray}
A_e(1996-98)		& = & 0.1572 \pm 0.0069 \pm 0.0027 
	\mbox{~~(from $e^+e^-$, $\mu^+\mu^-$ and $\tau^+\tau^-$);} \nonumber\\
A_{\mu}(1996-98)	& = & 0.147~~\pm 0.018~ \pm 0.002~
	\mbox{~~(from $\mu^+\mu^-$);~~and} \nonumber \\
A_{\tau}(1996-98)	& = & 0.127~~\pm 0.018~ \pm 0.004~
	\mbox{~~(from $\tau^+\tau^-$),} \nonumber
\end{eqnarray}
where the first error is statistical and second is due to systematic effects.
The numbers have been corrected for
the effect of backgrounds and the ``V-A effect'' for taus. 
The estimates for
$A_e$, $A_{\mu}$ and $A_{\tau}$ are obtained by fitting each lepton sample
separately by the maximum likelihood procedure. 
$A_e$ is obtained from all lepton species combined 
(combined from $A_e^e$, $A_e^{\mu}$ and $A_e^{\tau}$). 

Adding our published results from the 1993-95 data, our current best
estimates for leptonic asymmetry parameters at SLD are as follows:
\begin{eqnarray}
A_e(1993-98)		& = & 0.1558 \pm 0.0064
	\mbox{~~(from $e^+e^-$, $\mu^+\mu^-$ and $\tau^+\tau^-$);} \nonumber\\
A_{\mu}(1993-98)	& = & 0.137~~\pm 0.016~
	\mbox{~~(from $\mu^+\mu^-$);} \nonumber \\
A_{\tau}(1993-98)	& = & 0.142~~\pm 0.016~
	\mbox{~~(from $\tau^+\tau^-$);~~and} \nonumber \\
A_{l}(1993-98)		& = & 0.1523 \pm 0.0057, \nonumber
\end{eqnarray}
where statistical and systematic errors are combined.
The asymmetry parameters are consistent with lepton universality.
The global asymmetry parameter is referred to as $A_{l}$
(For $A_l$, systematic errors are conservatively taken to be fully correlated
between lepton species).

\section*{SUMMARY}

We report new direct measurements of the $Z^0$-lepton coupling 
asymmetry parameters $A_e$, $A_{\mu}$ and $A_{\tau}$, 
with polarized $Z^0$'s collected from 1993 through 1998 by the SLD detector 
at the SLAC Linear Collider.
Maximum likelihood fits to the reactions 
$e^-_{L,R} + e^+ \rightarrow Z^0 \rightarrow e^+e^-$, $\mu^+\mu^-$ 
and $\tau^+\tau^-$ are used to measure the parameters. 
The probability density function used in the fit incorporates all
three $s$-channel terms required from the tree-level calculations for the muon-
and tau-pair final states. 
The electron-pair final states are described by both
$s$- and $t$-channel $Z^0$ and photon exchange requiring ten cross section
terms, all of which are included in the probability density function. 
Whether three or ten terms, 
the probability density function used in the fit results
from convoluting the energy-dependent cross section formulas with a spectral
function.
The function incorporates initial state QED radiation, the intrinsic 
beam-energy spread and the effect of energy-dependent selection criteria. 
The parameters obtained from these fits require no further corrections 
for these effects. 
However, $A_{\tau}$ is corrected for a bias that results from the
V-A structure of tau decays and both tau- and  electron-pair events require
additional small corrections due to backgrounds. 
Preliminary results are summarized in the previous section.

Comparison of the $A_e$, $A_{\mu}$ and $A_{\tau}$ shows
no significant differences in these asymmetry parameters.
By assuming lepton universality, the weak mixing angle 
corresponding to the global asymmetry parameter, $A_{l}$, 
is given 
\begin{equation}
\sin^2\theta_W^{eff} = 0.23085 \pm 0.00073 .
\end{equation}
The weak mixing angle from $A_{LR}$ using hadrons yields~\cite{Abe:1997nj}
\begin{equation}
\sin^2\theta_W^{eff} = 0.23101 \pm 0.00029 .
\end{equation}
Those results are consistent and
the combined {\bf preliminary} result of the weak mixing angle at SLD is
\begin{equation}
\sin^2\theta_W^{eff} = 0.23099 \pm 0.00026 .
\end{equation}
Our result still differs from the most recent combined 
LEP I result~\cite{cite:LEPEWWG} by about $2.7\sigma$.
It is interesting to note that the LEP leptonic average
($\sin^2\theta_W^{eff}=0.23151 \pm 0.00033$), from the tau-polarization and 
the unpolarized forward-backward leptonic asymmetries, is consistent with our
result within $1.2\sigma$.
However the LEP hadronic average ($\sin^2\theta_W^{eff}=0.23230 \pm 0.00032$),
from the $b$-quark and $c$-quark unpolarized forward-backward asymmetries and 
the hadronic charge asymmetry,
is different from our result by $3.1\sigma$.

\section*{Acknowledgment}
We thank the personnel of the SLAC accelerator department and the
technical staffs of our collaborating institutions for their outstanding
efforts on our behalf. 
This work was supported by the Department of
Energy, the National Science Foundation, the Istituto Nazionale di
Fisica Nucleare of Italy, the Japan-US Cooperative Research Project
on High Energy Physics, and the Science and Engineering Research
Council of the United Kingdom.


\newpage
\section*{**List of Authors}

%
%
%
\begin{center}
\def\iADEL{$^{(1)}$}
\def\iAOMORI{$^{(2)}$}
\def\iBOLO{$^{(3)}$}
\def\iBRI{$^{(4)}$}
\def\iBRUN{$^{(5)}$}
\def\iBU{$^{(6)}$}
\def\iCINC{$^{(7)}$}
\def\iCOLO{$^{(8)}$}
\def\iCOLU{$^{(9)}$}
\def\iCSU{$^{(10)}$}
\def\iFERR{$^{(11)}$}
\def\iFRAS{$^{(12)}$}
\def\iILLI{$^{(13)}$}
\def\iJHU{$^{(14)}$}
\def\iLBL{$^{(15)}$}
\def\iLTU{$^{(16)}$}
\def\iMASS{$^{(17)}$}
\def\iMISSI{$^{(18)}$}
\def\iMIT{$^{(19)}$}
\def\iMOSCOW{$^{(20)}$}
\def\iNAGO{$^{(21)}$}
\def\iOREG{$^{(22)}$}
\def\iOXF{$^{(23)}$}
\def\iPADO{$^{(24)}$}
\def\iPERU{$^{(25)}$}
\def\iPISA{$^{(26)}$}
\def\iRAL{$^{(27)}$}
\def\iRUTG{$^{(28)}$}
\def\iSLAC{$^{(29)}$}
\def\iSOGA{$^{(30)}$}
\def\iSOONG{$^{(31)}$}
\def\iTENN{$^{(32)}$}
\def\iTOHO{$^{(33)}$}
\def\iUCSB{$^{(34)}$}
\def\iUCSC{$^{(35)}$}
\def\iUVIC{$^{(36)}$}
\def\iVAND{$^{(37)}$}
\def\iWASH{$^{(38)}$}
\def\iWISC{$^{(39)}$}
\def\iYALE{$^{(40)}$}

  \baselineskip=.75\baselineskip  
\mbox{Kenji  Abe\unskip,\iNAGO}
\mbox{Koya Abe\unskip,\iTOHO}
\mbox{T. Abe\unskip,\iSLAC}
\mbox{I. Adam\unskip,\iSLAC}
\mbox{T.  Akagi\unskip,\iSLAC}
\mbox{H. Akimoto\unskip,\iSLAC}
\mbox{N.J. Allen\unskip,\iBRUN}
\mbox{W.W. Ash\unskip,\iSLAC}
\mbox{D. Aston\unskip,\iSLAC}
\mbox{K.G. Baird\unskip,\iMASS}
\mbox{C. Baltay\unskip,\iYALE}
\mbox{H.R. Band\unskip,\iWISC}
\mbox{M.B. Barakat\unskip,\iLTU}
\mbox{O. Bardon\unskip,\iMIT}
\mbox{T.L. Barklow\unskip,\iSLAC}
\mbox{G.L. Bashindzhagyan\unskip,\iMOSCOW}
\mbox{J.M. Bauer\unskip,\iMISSI}
\mbox{G. Bellodi\unskip,\iOXF}
\mbox{A.C. Benvenuti\unskip,\iBOLO}
\mbox{G.M. Bilei\unskip,\iPERU}
\mbox{D. Bisello\unskip,\iPADO}
\mbox{G. Blaylock\unskip,\iMASS}
\mbox{J.R. Bogart\unskip,\iSLAC}
\mbox{G.R. Bower\unskip,\iSLAC}
\mbox{J.E. Brau\unskip,\iOREG}
\mbox{M. Breidenbach\unskip,\iSLAC}
\mbox{W.M. Bugg\unskip,\iTENN}
\mbox{D. Burke\unskip,\iSLAC}
\mbox{T.H. Burnett\unskip,\iWASH}
\mbox{P.N. Burrows\unskip,\iOXF}
\mbox{R.M. Byrne\unskip,\iMIT}
\mbox{A. Calcaterra\unskip,\iFRAS}
\mbox{D. Calloway\unskip,\iSLAC}
\mbox{B. Camanzi\unskip,\iFERR}
\mbox{M. Carpinelli\unskip,\iPISA}
\mbox{R. Cassell\unskip,\iSLAC}
\mbox{R. Castaldi\unskip,\iPISA}
\mbox{A. Castro\unskip,\iPADO}
\mbox{M. Cavalli-Sforza\unskip,\iUCSC}
\mbox{A. Chou\unskip,\iSLAC}
\mbox{E. Church\unskip,\iWASH}
\mbox{H.O. Cohn\unskip,\iTENN}
\mbox{J.A. Coller\unskip,\iBU}
\mbox{M.R. Convery\unskip,\iSLAC}
\mbox{V. Cook\unskip,\iWASH}
\mbox{R.F. Cowan\unskip,\iMIT}
\mbox{D.G. Coyne\unskip,\iUCSC}
\mbox{G. Crawford\unskip,\iSLAC}
\mbox{C.J.S. Damerell\unskip,\iRAL}
\mbox{M.N. Danielson\unskip,\iCOLO}
\mbox{M. Daoudi\unskip,\iSLAC}
\mbox{N. de Groot\unskip,\iBRI}
\mbox{R. Dell'Orso\unskip,\iPERU}
\mbox{P.J. Dervan\unskip,\iBRUN}
\mbox{R. de Sangro\unskip,\iFRAS}
\mbox{M. Dima\unskip,\iCSU}
\mbox{D.N. Dong\unskip,\iMIT}
\mbox{M. Doser\unskip,\iSLAC}
\mbox{R. Dubois\unskip,\iSLAC}
\mbox{B.I. Eisenstein\unskip,\iILLI}
\mbox{I.Erofeeva\unskip,\iMOSCOW}
\mbox{V. Eschenburg\unskip,\iMISSI}
\mbox{E. Etzion\unskip,\iWISC}
\mbox{S. Fahey\unskip,\iCOLO}
\mbox{D. Falciai\unskip,\iFRAS}
\mbox{C. Fan\unskip,\iCOLO}
\mbox{J.P. Fernandez\unskip,\iUCSC}
\mbox{M.J. Fero\unskip,\iMIT}
\mbox{K. Flood\unskip,\iMASS}
\mbox{R. Frey\unskip,\iOREG}
\mbox{J. Gifford\unskip,\iUVIC}
\mbox{T. Gillman\unskip,\iRAL}
\mbox{G. Gladding\unskip,\iILLI}
\mbox{S. Gonzalez\unskip,\iMIT}
\mbox{E.R. Goodman\unskip,\iCOLO}
\mbox{E.L. Hart\unskip,\iTENN}
\mbox{J.L. Harton\unskip,\iCSU}
\mbox{K. Hasuko\unskip,\iTOHO}
\mbox{S.J. Hedges\unskip,\iBU}
\mbox{S.S. Hertzbach\unskip,\iMASS}
\mbox{M.D. Hildreth\unskip,\iSLAC}
\mbox{J. Huber\unskip,\iOREG}
\mbox{M.E. Huffer\unskip,\iSLAC}
\mbox{E.W. Hughes\unskip,\iSLAC}
\mbox{X. Huynh\unskip,\iSLAC}
\mbox{H. Hwang\unskip,\iOREG}
\mbox{M. Iwasaki\unskip,\iOREG}
\mbox{D.J. Jackson\unskip,\iRAL}
\mbox{P. Jacques\unskip,\iRUTG}
\mbox{J.A. Jaros\unskip,\iSLAC}
\mbox{Z.Y. Jiang\unskip,\iSLAC}
\mbox{A.S. Johnson\unskip,\iSLAC}
\mbox{J.R. Johnson\unskip,\iWISC}
\mbox{R.A. Johnson\unskip,\iCINC}
\mbox{T. Junk\unskip,\iSLAC}
\mbox{R. Kajikawa\unskip,\iNAGO}
\mbox{M. Kalelkar\unskip,\iRUTG}
\mbox{Y. Kamyshkov\unskip,\iTENN}
\mbox{H.J. Kang\unskip,\iRUTG}
\mbox{I. Karliner\unskip,\iILLI}
\mbox{H. Kawahara\unskip,\iSLAC}
\mbox{Y.D. Kim\unskip,\iSOGA}
\mbox{M.E. King\unskip,\iSLAC}
\mbox{R. King\unskip,\iSLAC}
\mbox{R.R. Kofler\unskip,\iMASS}
\mbox{N.M. Krishna\unskip,\iCOLO}
\mbox{R.S. Kroeger\unskip,\iMISSI}
\mbox{M. Langston\unskip,\iOREG}
\mbox{A. Lath\unskip,\iMIT}
\mbox{D.W.G. Leith\unskip,\iSLAC}
\mbox{V. Lia\unskip,\iMIT}
\mbox{C.Lin\unskip,\iMASS}
\mbox{M.X. Liu\unskip,\iYALE}
\mbox{X. Liu\unskip,\iUCSC}
\mbox{M. Loreti\unskip,\iPADO}
\mbox{A. Lu\unskip,\iUCSB}
\mbox{H.L. Lynch\unskip,\iSLAC}
\mbox{J. Ma\unskip,\iWASH}
\mbox{M. Mahjouri\unskip,\iMIT}
\mbox{G. Mancinelli\unskip,\iRUTG}
\mbox{S. Manly\unskip,\iYALE}
\mbox{G. Mantovani\unskip,\iPERU}
\mbox{T.W. Markiewicz\unskip,\iSLAC}
\mbox{T. Maruyama\unskip,\iSLAC}
\mbox{H. Masuda\unskip,\iSLAC}
\mbox{E. Mazzucato\unskip,\iFERR}
\mbox{A.K. McKemey\unskip,\iBRUN}
\mbox{B.T. Meadows\unskip,\iCINC}
\mbox{G. Menegatti\unskip,\iFERR}
\mbox{R. Messner\unskip,\iSLAC}
\mbox{P.M. Mockett\unskip,\iWASH}
\mbox{K.C. Moffeit\unskip,\iSLAC}
\mbox{T.B. Moore\unskip,\iYALE}
\mbox{M.Morii\unskip,\iSLAC}
\mbox{D. Muller\unskip,\iSLAC}
\mbox{V. Murzin\unskip,\iMOSCOW}
\mbox{T. Nagamine\unskip,\iTOHO}
\mbox{S. Narita\unskip,\iTOHO}
\mbox{U. Nauenberg\unskip,\iCOLO}
\mbox{H. Neal\unskip,\iSLAC}
\mbox{M. Nussbaum\unskip,\iCINC}
\mbox{N. Oishi\unskip,\iNAGO}
\mbox{D. Onoprienko\unskip,\iTENN}
\mbox{L.S. Osborne\unskip,\iMIT}
\mbox{R.S. Panvini\unskip,\iVAND}
\mbox{C.H. Park\unskip,\iSOONG}
\mbox{T.J. Pavel\unskip,\iSLAC}
\mbox{I. Peruzzi\unskip,\iFRAS}
\mbox{M. Piccolo\unskip,\iFRAS}
\mbox{L. Piemontese\unskip,\iFERR}
\mbox{K.T. Pitts\unskip,\iOREG}
\mbox{R.J. Plano\unskip,\iRUTG}
\mbox{R. Prepost\unskip,\iWISC}
\mbox{C.Y. Prescott\unskip,\iSLAC}
\mbox{G.D. Punkar\unskip,\iSLAC}
\mbox{J. Quigley\unskip,\iMIT}
\mbox{B.N. Ratcliff\unskip,\iSLAC}
\mbox{T.W. Reeves\unskip,\iVAND}
\mbox{J. Reidy\unskip,\iMISSI}
\mbox{P.L. Reinertsen\unskip,\iUCSC}
\mbox{P.E. Rensing\unskip,\iSLAC}
\mbox{L.S. Rochester\unskip,\iSLAC}
\mbox{P.C. Rowson\unskip,\iCOLU}
\mbox{J.J. Russell\unskip,\iSLAC}
\mbox{O.H. Saxton\unskip,\iSLAC}
\mbox{T. Schalk\unskip,\iUCSC}
\mbox{R.H. Schindler\unskip,\iSLAC}
\mbox{B.A. Schumm\unskip,\iUCSC}
\mbox{J. Schwiening\unskip,\iSLAC}
\mbox{S. Sen\unskip,\iYALE}
\mbox{V.V. Serbo\unskip,\iSLAC}
\mbox{M.H. Shaevitz\unskip,\iCOLU}
\mbox{J.T. Shank\unskip,\iBU}
\mbox{G. Shapiro\unskip,\iLBL}
\mbox{D.J. Sherden\unskip,\iSLAC}
\mbox{K.D. Shmakov\unskip,\iTENN}
\mbox{C. Simopoulos\unskip,\iSLAC}
\mbox{N.B. Sinev\unskip,\iOREG}
\mbox{S.R. Smith\unskip,\iSLAC}
\mbox{M.B. Smy\unskip,\iCSU}
\mbox{J.A. Snyder\unskip,\iYALE}
\mbox{H. Staengle\unskip,\iCSU}
\mbox{A. Stahl\unskip,\iSLAC}
\mbox{P. Stamer\unskip,\iRUTG}
\mbox{H. Steiner\unskip,\iLBL}
\mbox{R. Steiner\unskip,\iADEL}
\mbox{M.G. Strauss\unskip,\iMASS}
\mbox{D. Su\unskip,\iSLAC}
\mbox{F. Suekane\unskip,\iTOHO}
\mbox{A. Sugiyama\unskip,\iNAGO}
\mbox{S. Suzuki\unskip,\iNAGO}
\mbox{M. Swartz\unskip,\iJHU}
\mbox{A. Szumilo\unskip,\iWASH}
\mbox{T. Takahashi\unskip,\iSLAC}
\mbox{F.E. Taylor\unskip,\iMIT}
\mbox{J. Thom\unskip,\iSLAC}
\mbox{E. Torrence\unskip,\iMIT}
\mbox{N.K. Toumbas\unskip,\iSLAC}
\mbox{T. Usher\unskip,\iSLAC}
\mbox{C. Vannini\unskip,\iPISA}
\mbox{J. Va'vra\unskip,\iSLAC}
\mbox{E. Vella\unskip,\iSLAC}
\mbox{J.P. Venuti\unskip,\iVAND}
\mbox{R. Verdier\unskip,\iMIT}
\mbox{P.G. Verdini\unskip,\iPISA}
\mbox{D.L. Wagner\unskip,\iCOLO}
\mbox{S.R. Wagner\unskip,\iSLAC}
\mbox{A.P. Waite\unskip,\iSLAC}
\mbox{S. Walston\unskip,\iOREG}
\mbox{S.J. Watts\unskip,\iBRUN}
\mbox{A.W. Weidemann\unskip,\iTENN}
\mbox{E. R. Weiss\unskip,\iWASH}
\mbox{J.S. Whitaker\unskip,\iBU}
\mbox{S.L. White\unskip,\iTENN}
\mbox{F.J. Wickens\unskip,\iRAL}
\mbox{B. Williams\unskip,\iCOLO}
\mbox{D.C. Williams\unskip,\iMIT}
\mbox{S.H. Williams\unskip,\iSLAC}
\mbox{S. Willocq\unskip,\iMASS}
\mbox{R.J. Wilson\unskip,\iCSU}
\mbox{W.J. Wisniewski\unskip,\iSLAC}
\mbox{J. L. Wittlin\unskip,\iMASS}
\mbox{M. Woods\unskip,\iSLAC}
\mbox{G.B. Word\unskip,\iVAND}
\mbox{T.R. Wright\unskip,\iWISC}
\mbox{J. Wyss\unskip,\iPADO}
\mbox{R.K. Yamamoto\unskip,\iMIT}
\mbox{J.M. Yamartino\unskip,\iMIT}
\mbox{X. Yang\unskip,\iOREG}
\mbox{J. Yashima\unskip,\iTOHO}
\mbox{S.J. Yellin\unskip,\iUCSB}
\mbox{C.C. Young\unskip,\iSLAC}
\mbox{H. Yuta\unskip,\iAOMORI}
\mbox{G. Zapalac\unskip,\iWISC}
\mbox{R.W. Zdarko\unskip,\iSLAC}
\mbox{J. Zhou\unskip.\iOREG}

\it
  \vskip \baselineskip                   
  \centerline{(The SLD Collaboration)}   
  \vskip \baselineskip        
  \baselineskip=.75\baselineskip   
\iADEL
  Adelphi University, Garden City, New York 11530, \break
\iAOMORI
  Aomori University, Aomori , 030 Japan, \break
\iBOLO
  INFN Sezione di Bologna, I-40126, Bologna, Italy, \break
\iBRI
  University of Bristol, Bristol, U.K., \break
\iBRUN
  Brunel University, Uxbridge, Middlesex, UB8 3PH United Kingdom, \break
\iBU
  Boston University, Boston, Massachusetts 02215, \break
\iCINC
  University of Cincinnati, Cincinnati, Ohio 45221, \break
\iCOLO
  University of Colorado, Boulder, Colorado 80309, \break
\iCOLU
  Columbia University, New York, New York 10533, \break
\iCSU
  Colorado State University, Ft. Collins, Colorado 80523, \break
\iFERR
  INFN Sezione di Ferrara and Universita di Ferrara, I-44100 Ferrara, Italy, \break
\iFRAS
  INFN Lab. Nazionali di Frascati, I-00044 Frascati, Italy, \break
\iILLI
  University of Illinois, Urbana, Illinois 61801, \break
\iJHU
  Johns Hopkins University,  Baltimore, Maryland 21218-2686, \break
\iLBL
  Lawrence Berkeley Laboratory, University of California, Berkeley, California 94720, \break
\iLTU
  Louisiana Technical University, Ruston,Louisiana 71272, \break
\iMASS
  University of Massachusetts, Amherst, Massachusetts 01003, \break
\iMISSI
  University of Mississippi, University, Mississippi 38677, \break
\iMIT
  Massachusetts Institute of Technology, Cambridge, Massachusetts 02139, \break
\iMOSCOW
  Institute of Nuclear Physics, Moscow State University, 119899, Moscow Russia, \break
\iNAGO
  Nagoya University, Chikusa-ku, Nagoya, 464 Japan, \break
\iOREG
  University of Oregon, Eugene, Oregon 97403, \break
\iOXF
  Oxford University, Oxford, OX1 3RH, United Kingdom, \break
\iPADO
  INFN Sezione di Padova and Universita di Padova I-35100, Padova, Italy, \break
\iPERU
  INFN Sezione di Perugia and Universita di Perugia, I-06100 Perugia, Italy, \break
\iPISA
  INFN Sezione di Pisa and Universita di Pisa, I-56010 Pisa, Italy, \break
\iRAL
  Rutherford Appleton Laboratory, Chilton, Didcot, Oxon OX11 0QX United Kingdom, \break
\iRUTG
  Rutgers University, Piscataway, New Jersey 08855, \break
\iSLAC
  Stanford Linear Accelerator Center, Stanford University, Stanford, California 94309, \break
\iSOGA
  Sogang University, Seoul, Korea, \break
\iSOONG
  Soongsil University, Seoul, Korea 156-743, \break
\iTENN
  University of Tennessee, Knoxville, Tennessee 37996, \break
\iTOHO
  Tohoku University, Sendai 980, Japan, \break
\iUCSB
  University of California at Santa Barbara, Santa Barbara, California 93106, \break
\iUCSC
  University of California at Santa Cruz, Santa Cruz, California 95064, \break
\iUVIC
  University of Victoria, Victoria, British Columbia, Canada V8W 3P6, \break
\iVAND
  Vanderbilt University, Nashville,Tennessee 37235, \break
\iWASH
  University of Washington, Seattle, Washington 98105, \break
\iWISC
  University of Wisconsin, Madison,Wisconsin 53706, \break
\iYALE
  Yale University, New Haven, Connecticut 06511. \break

\rm
%

\end{center}

\begin{thebibliography}{9}  

\bibitem{Erler:1998ig}
See, for example,
J.~Erler and P.~Langacker,
``Electroweak Model and Constraints on New Physics,''
Eur. Phys. J. {\bf C3}, 90 (1998).

\bibitem{Abbaneo:1999ub}
D.~Abbaneo {\it et al.}
[LEP Collaboration],
``A Combination of preliminary electroweak measurements and constraints on
                  the standard model,''
CERN-EP-99-015.

\bibitem{Abe:1997xm}
K.~Abe {\it et al.}
[SLD Collaboration],
``Direct measurement of leptonic coupling asymmetries with polarized Zs,''
Phys. Rev. Lett. {\bf 79}, 804 (1997)
hep-ex/9704012;\\
M.B.~Smy,
``Measurement of Z0 lepton coupling asymmetries,''
SLAC-R-0515, PhD Thesis, Colorado State University, July 1997.

\bibitem{Abe:1997nj}
K.~Abe {\it et al.}
[SLD Collaboration],
``An improved measurement of the left-right Z0 cross-section asymmetry,''
Phys. Rev. Lett. {\bf 78}, 2075 (1997)
hep-ex/9611011;\\
The latest result was presented by
J.~Brau [SLD Collaboration], talk presented at 
the International Europhysics Conference on High Energy Physics,
15-21 July 1999, Tampere, Finland


\bibitem{Woods:1996ph}
M.~Woods,
``The Polarized electron beam for the SLAC linear collider,''
{\it  In *Amsterdam 1996, 12th International Symposium on High-energy 
Spin Physics (SPIN 96)* 623-627}
hep-ex/9611006.

\bibitem{Woods:1996nz}
M.~Woods
[SLD Collaboration],
``The Scanning Compton polarimeter for the SLD experiment,''
{\it  In *Amsterdam 1996, 12th International Symposium on High-energy 
Spin Physics (SPIN 96)* 843-845}
hep-ex/9611005.

\bibitem{unknown:1984rp}
[SLD Collaboration],
``SLD Design Report,''
SLAC-0273, 1984.

\bibitem{Abe:1997bu}
K.~Abe {\it et al.},
``Design and performance of the SLD vertex detector, a 307 Mpixel tracking
                  system,''
Nucl. Instrum. Meth. {\bf A400}, 287 (1997);\\
N.B.~Sinev {\it et al.}
[SLD Collaboration],
``Performance of the new vertex detector at SLD,''
IEEE Trans. Nucl. Sci. {\bf 44}, 587 (1997);\\
J.E.~Brau
[SLD Collaboration],
``Design and performance of the new CCD vertex detector at SLD and
                  implications for the next linear collider,''
Nucl. Instrum. Meth. {\bf A418}, 52 (1998).

\bibitem{VERTEX99byT.Abe}
T.~Abe [SLD Collaboration], talk presented at 
8th International Workshop on Vertex Detectors,
20-25 June 1999, Texel, Netherlands

\bibitem{cite:gterm} 
The photon $s$-channel cross section terms are
$\displaystyle\left(\frac{d\sigma}{dx}\right)_{\gamma(s)}\propto(1+x^2)$
and
$\displaystyle\left(\frac{d\sigma}{dx}\right)_{\gamma(s)Z(s)}\propto
\frac{v_e}{a_e}\frac{v_l}{a_l}\times
\left(\left(1-P\frac{a_e}{v_e}\right)\left(1+x^2\right)+
\left(\frac{a_e}{v_e}-P\right)\frac{a_l}{v_l}2x\right)$.
The dominant $t$-channel cross section for electrons is
$\displaystyle\left(\frac{d\sigma}{dx}\right)_{\gamma(t)}\propto
\frac{4+(x+1)^2}{(x-1)^2}$.

\bibitem{cite:polsign}
We define the polarization as $P=(R-L)/(R+L)$ where $L$($R$) is the number of
left-handed (right-handed) electrons.

\bibitem{Martinez:1991ta}
M.~Martinez, L.~Garrido, R.~Miquel, J.L.~Harton and R.~Tanaka,
``Model independent fitting to the Z line shape,''
Z. Phys. {\bf C49}, 645 (1991).

\bibitem{Martinez:1992wy}
M.~Martinez and R.~Miquel,
``Fitting the e+ e- $\to$ e+ e- line shape,''
Z. Phys. {\bf C53}, 115 (1992).

\bibitem{ZSCAN}
P.~Rowson, R.~Frey, S.~Hertzbach, R.~Kofler, M.~Swartz and M.~Woods,
``Calibration of the WISRD Energy Spectrometer with a Z Peak Scan,''
SLD-Note-264 (1999).

\bibitem{Field:1998sh}
R.C.~Field, M.~Woods, J.~Zhou, R.~Frey and A.~Arodzero,
``Measurement of electron beam polarization from the energy asymmetry of
                  Compton scattered photons,''
IEEE Trans. Nucl. Sci. {\bf 45}, 670 (1998).

\bibitem{Berridge:1997pw}
S.C.~Berridge {\it et al.},
``Quartz fiber/tungsten calorimeter for Compton polarimeter at SLD,''
{\it  In *Tucson 1997, Calorimetry in high energy physics* 170-181}.

\bibitem{cite:polmeasurement}
M.~Fero {\it et al.},
SLD-Physics-Note-50 (1996), to be updated.

\bibitem{cite:finalpol}
The latest study by the polarimeter group ~\cite{Abe:1997nj}
estimates the errors are $\delta P/P=$0.67\% and 0.5\% for 1996 and 1997-98, 
respectively, which are very close to the final numbers.
In this analysis, the preliminary numbers are chosen to estimate
the uncertainty conservatively.

\bibitem{cite:UncertaintyofECM}
The uncertainty was obtained by using the center-of-mass energy measured by 
the WISRD energy spectrometer for events passing the pre-selection cuts.
The obtained uncertainty is higher than the value estimated in 
Ref.~\cite{ZSCAN}.
We use the uncertainty of 50 MeV to estimate the systematic error 
conservatively for now.

\bibitem{Tsai:1971vv}
Y.~Tsai,
``Decay Correlations Of Heavy Leptons In E+ E- $\to$ Lepton+ Lepton-,''
Phys. Rev. {\bf D4}, 2821 (1971).

\bibitem{cite:LEPEWWG}
The latest LEP-electroweak results are summarized in 
the LEP Electroweak Working Group Home Page 
(http://www.cern.ch/LEPEWWG/Welcome.html).
\end{thebibliography}
\enddocument